# Nonlinear two-dimensional terahertz photon echo and rotational spectroscopy in the gas phase

Jian Lu[1], Yaqing Zhang[1], Harold Y. Hwang[1], Benjamin K. Ofori-Okai[1], Sharly Fleischer[2] and Keith A. Nelson[1, *]

[1]Department of Chemistry, Massachusetts Institute of Technology, Cambridge, Massachusetts 02139, USA.

[2]Department of Chemical Physics, Tel-Aviv University, Tel Aviv 69978, Israel.

[*]Email: kanelson@mit.edu

**Ultrafast two-dimensional spectroscopy utilizes correlated multiple light-matter interactions for retrieving dynamic features that may otherwise be hidden under the linear spectrum. Its extension to the terahertz regime of the electromagnetic spectrum, where a rich variety of material degrees of freedom reside, remains an experimental challenge. Here we report ultrafast two-dimensional terahertz spectroscopy of gas-phase molecular rotors at room temperature. Using time-delayed terahertz pulse pairs, we observe photon echoes and other nonlinear signals resulting from molecular dipole orientation induced by three terahertz field-dipole interactions. The nonlinear time-domain orientation signals are mapped into the frequency domain in two-dimensional rotational spectra which reveal J-state-resolved nonlinear rotational dynamics. The approach enables direct observation of correlated rotational transitions and may reveal rotational coupling and relaxation pathways in the ground electronic and vibrational state.**

Recent years have witnessed increasing interest in two-dimensional infrared (2D IR) vibrational spectroscopy techniques for studying structural dynamics and correlations between coupled molecular motions in biological systems such as water[1], proteins[2] and DNA[3]. Multidimensional optical spectroscopies were applied to probe the high-order correlations of excitons in quantum wells[4] and organic complexes[5]. Yet it remains an experimental challenge to extend multidimensional spectroscopies into the terahertz (THz) frequency range where a rich variety of material degrees of freedom including gas molecular rotations, lattice vibrations in solids, magnetization dynamics in magnetically ordered materials and many others[6] find fundamental and technological importance. Despite challenges involved, there have been examples of 2D THz, 2D Raman and 2D THz-Raman spectroscopies in the studies of electronic nonlinearities in solids[7], coupled Raman-active vibrational modes in liquid molecules[8], and hydrogen bond dynamics in water[9].

Rotations of gas molecules have been the subjects of intensive recent efforts in coherent spectroscopy and coherent control, motivated by interests in rotational angular momentum and energy relaxation processes, high-order optical interactions with multi-level quantum systems, and applications of molecular alignment and orientation for various forms of orbital tomography, for example using high-harmonic generation to probe the molecular orbital structures of rotating molecules[10, 11]. Strong optical fields have been used to drive high-order rotational coherences and to populate high-lying rotational levels[12], to control molecular alignment dynamics[13], and to induce rotational alignment echoes[14, 15]. Strong microwave fields can drive high-order rotational responses and have been used for double-resonance and 2D rotational spectroscopies[16-21] mostly at low rotational temperatures. In recent work, we demonstrated second-order interactions between terahertz-frequency (THz) fields and molecular rotations, measuring net molecular dipole orientation, manipulating 2-quantum coherences (2QC), and observing rotational populations and their superradiant decays[22-24]. Higher-order THz field interactions have been treated theoretically, including photon echoes resulting from molecular orientation predicted by classical phase-space distribution calculations[14] and full quantum mechanical simulations[25]. The recent experimental and theoretical results motivate developments in nonlinear multidimensional rotational spectroscopy in the THz frequency range that includes many of the transitions of small molecules at ordinary temperatures.

In this work we used variably delayed single-cycle THz pulse pairs to demonstrate the full suite of third-order rotational responses – rephasing (R, i.e. photon echo), non-rephasing (NR), 2-quantum (2Q), and pump-probe (PP) signals – separated from each other and fully $J$-state-resolved through 2D THz time-domain spectroscopy measurements. The results demonstrate the capability for full elucidation of rotational population and coherence dynamics with sub-picosecond time resolution, applicable to gases under conditions that could lead to extremely fast relaxation and dephasing such as high (including supercritical) pressure and temperature, flames, and reactive mixtures. We also observe higher-order responses due to five field-dipole interactions, showing that extensive THz field control over successive populations and coherences can be executed in order to explore microscopic pathways involving multiple rotational $J$ levels.

An ultrashort THz pulse induces coherent molecular rotations via resonant field-dipole interactions. The linear interaction term, $H_1 = -\vec{\mu} \cdot \vec{E}_{\text{THz}} = -\mu E_{\text{THz}} \cos\theta$, where $\theta$ is the angle between molecular dipoles and the THz electric field polarization, couples adjacent rotational states $J$ and $J+1$ yielding 1-quantum coherences (1QCs) among the thermally populated rotational states. Each 1QC oscillates at a frequency of $f_{J,J+1} = 2Bc(J+1)$ where $B$ is the rotational constant in cm$^{-1}$ units and $c$ the speed of light. The superposition of 1QCs with frequencies that are integer multiples of the lowest frequency $2Bc$ gives rise to short duration events of net molecular dipole orientation with a nonzero orientation factor $\langle\cos\theta\rangle$ that recur with the quantum rotational revival[26] period of $T_{\text{rev}} = (2Bc)^{-1}$. This is analogous to the laser pulse train from a mode-locked oscillator with a distribution of equally spaced cavity modes[27]. Upon each revival, the net dipole orientation results in a macroscopic polarization that emits a burst of coherent THz-frequency radiation. The free-induction-decay (FID) signal thus consists of a sequence of such bursts separated by $T_{\text{rev}}$. Two successive interactions between the incident THz field and the molecular dipoles can couple rotational states $J$ and $J+2$ to yield 2QCs whose superposition shows dipole alignment revivals with the period equal to $T_{\text{rev}}/2$, and can transfer population between states $J$ and $J+1$, both outcomes manifested as a transient birefringence of the molecular ensemble[22]. It has been shown that two time-delayed field-dipole interactions can coherently control the rotational populations and coherences to significantly enhance the 2QC amplitudes and the corresponding degree of net molecular alignment $\langle\cos^2\theta\rangle$ [23]. These examples of THz second-order interactions and spectroscopy set the stage for multidimensional spectroscopy at third and higher order.

**Results**

**Experimental schematic and sample linear orientation response.** The experimental setup is shown schematically in Fig. 1a. Two THz pulses denoted as A and B separated by a time delay $\tau$ were focused into a static pressure sample cell in collinear geometry. The transmitted THz fields and the subsequent emitted signals were detected through the birefringence they induce in an electro-optic (EO) crystal, measured using a 100 fs optical readout pulse (EO sampling[28]) that was delayed by time $t$ relative to pulse B. The measurements were performed on gaseous acetonitrile (CH$_3$CN, dipole moment $\mu = 3.92$ debye) at room temperature and 70 torr pressure. The rotational constant is $B = 0.310$ cm$^{-1}$, corresponding to $T_{\text{rev}} = 54.5$ ps. A single THz excitation pulse was used to measure the FID signal from one field-dipole interaction, which shows the expected revivals (Fig. 1b, blue dashed boxes). The rotational transitions from thermally populated $J$ levels in CH$_3$CN overlap with the spectrum of our THz field as shown in Fig. 1c. Fourier transformation of the FID signal in Fig. 1b reveals the rotational transition peaks within our THz excitation bandwidth, as shown in Fig. 1d.

**THz photon echoes and other nonlinear signals from molecular orientation.** We implemented a differential chopping detection method[29] to extract the nonlinear signals resulting from two THz pulses interacting with the molecular dipoles (see Supplementary Information for details). There are four distinct types of signals observed that arise from three THz-field interactions with the molecular dipoles, i.e. four third-order signal types that are all distinguished in the time-domain traces of $E_{\text{NL}}(\tau,t)$:

(1) R (i.e. photon echo) and (2) NR signals, generated through one field interaction with pulse A and, after time delay $\tau$, two field interactions with pulse B;

(3) 2Q and (4) PP signals, arising from two instantaneous interactions with pulse A and, after time delay $\tau$, one interaction with pulse B.

These signals are of the same origins as those in 2D IR and optical spectroscopies. The nonlinear traces $E_{\mathrm{NL}}(\tau,t)$ at varying delays $\tau > 0$ (color coded in figure legend) are shown in Figs. 2a and 2b for selected time delays $\tau$. Since all the signals are emitted from the full collection of third-order 1QCs involving the entire set of thermally populated $J$ levels whose transition frequencies are included in the THz pulse spectrum, they all take the same form as the first-order FID, namely periodic bursts of THz radiation separated by revival time $T_{\mathrm{rev}}$. However, the signals do not all appear at the same time after the two incident THz pulses. For any delay $\tau < T_{\mathrm{rev}}$ between pulses A and B, the first R (photon echo) signal (1) appears only after an equal time $t = \tau$ after pulse B, and including their revivals the echo signals appear at $t = \tau + nT_{\mathrm{rev}}$ ($n = 0, 1, 2, 3 \ldots$). The NR signals (2) appear as changes to the first-order FID amplitude from pulse A caused by the action of pulse B, so they appear at times $t = -\tau + nT_{\mathrm{rev}}$ ($n = 1, 2, 3 \ldots$). The 2Q signals (3) appear at $t = -2\tau + nT_{\mathrm{rev}}$ ($n = 1, 2, 3$) as the 2QCs induced by pulse A evolve with approximately twice the frequency of 1QCs for time $\tau$, after which they are projected by pulse B back to the 1QC manifold and continue to evolve with the 1QC frequency. Thus, the completion of the third-order coherence rotational cycle occurs $2\tau$ earlier than the 1QC revivals of pulse B (i.e. $\tau$ earlier than the 1QC revivals of pulse A). The PP signals (4) appear coincident with pulse B whose absorption is modified by the nonthermal population distribution created by pulse A, and including their revivals the signals appear at times $t = nT_{\mathrm{rev}}$ ($n = 0, 1, 2, 3 \ldots$). Due to their small amplitudes, the 2Q and PP time-domain signals are shown in detail in the Supplementary Information. At $2T_{\mathrm{rev}} > \tau > T_{\mathrm{rev}}$, there is an additional R signal arising from the 1QC revival before the arrival of pulse B, i.e. the echo signals appears at $t = \tau + nT_{\mathrm{rev}}$ ($n = -1, 0, 1, 2 \ldots$). Generalizing the description above, the times at which the various third-order polarizations generating our nonlinear signals appear are given not only by the arrival times of pulses A and B, but by all the first-order polarizations at each revival that the incident pulses induce, i.e. each revival can contribute to the nonlinear signals equivalently to the incident pulse that induced it. This is elaborated further in the Supplementary Information.

The experimental traces were compared with semiclassical calculations of the dipole orientation responses in Figs. 2c and 2d. The simulations were performed by numerically integrating the Liouville-von Neumann equation of the density matrix under decoherence- and decay-free conditions (see Supplementary Information for simulation methods). The time-derivative of the ensemble-averaged orientation factor, $d\cos\langle\theta\rangle_{\mathrm{NL}}/dt$, was calculated to account for the THz field radiated from the collective polarization formed by the dipole orientation and showed good agreement with the experimental data. A noticeable difference is that the simulated photon echo signal peaks at inter-pulse delay time $\tau = T_{\mathrm{rev}}/2$ while in the experimental data the maximum occurs earlier. This is attributed to the relatively fast decoherence of rotations under our experimental pressure and temperature conditions.

**2D THz rotational spectra.** 2D time-domain nonlinear signal $E_{\mathrm{NL}}(\tau,t)$ was recorded as a function of $\tau$ and $t$ (see Supplementary Information). A numerical 2D Fourier transformation with respect to $\tau$ and $t$ yielded the complex 2D rotational spectrum as a function of the corresponding frequency variables $\nu$ and $f$. The 2D magnitude spectrum is shown in Fig. 3. All of the signals are fully $J$-resolved in both dimensions. The full 2D spectrum can be separated into R and NR quadrants, as the R signals are distinguished from the NR signals by negative values of the excitation frequency $\nu$ due to the reversed phase accumulation of the former. The 2D spectrum clearly includes all four of the signal types discussed above. Examples of each signal type are discussed as follows. For the R and NR signals, THz pulse A interacts once with the molecular dipoles to produce first-order 1QCs described by density matrix terms $|J\rangle\langle J+1|$. After inter-pulse delay $\tau$, pulse B interacts twice with the dipoles, first to produce rotational populations $|J+1\rangle\langle J+1|$ and from these to generate third-order 1QCs $|J\rangle\langle J+1|$ (NR) or $|J+1\rangle\langle J|$ (R) from which the measured signals are radiated during time $t$. The third-order nonlinear signal field $E_{NL}(\tau,t)$ shows oscillations at the 1QC frequencies along the two time variables, and in this example the frequencies are the same so Fourier transformation of the signal with respect to both time variables yields $J$-state-resolved diagonal peaks in the 2D spectrum. For 2Q and PP signals, pulse A interacts twice with the molecular dipoles to produce either 2QCs $|J\rangle\langle J+2|$ or populations $|J+1\rangle\langle J+1|$ respectively. After inter-pulse delay $\tau$, pulse B interacts once with the dipoles to produce third-order 1QCs $|J\rangle\langle J+1|$ that radiate the measured signals during time $t$. The 2Q signal field $E_{NL}(\tau,t)$ shows oscillations as a function of $\tau$ at the 2QC frequencies and oscillations as a function of $t$ at the 1QC frequencies, giving rise to $J$-state-resolved peaks at $\nu \cong 2f$. For PP signals, there is no coherence evolution during the inter-pulse delay so the signal appears in the 2D spectrum at zero frequency along $\nu$ and at $J$-state-resolved positions along $f$.

The NR and R quadrants are plotted separately in Figs. 4a and 4b, with the R quadrant $f$ frequency values made positive. $J$-state-resolved diagonal peaks are clearly observed along $\nu = f$ in both plots, consistent with the discussion above. In addition, for each excitation frequency $\nu = f_{J,J+1} = 2Bc(J+1)$, off-diagonal peaks are observed at detection frequencies $f = f_{J-1,J} = 2BcJ$ and $f = f_{J+1,J+2} = 2Bc(J+2)$, indicating that the first-order $|J\rangle\langle J+1|$ coherences induced via one interaction of the molecular dipoles with THz pulse A are correlated via two interactions with THz pulse B with not only the third-order $|J\rangle\langle J+1|$ coherences discussed above but also the third-order coherences $|J-1\rangle\langle J|$ and $|J+1\rangle\langle J+2|$ involving two neighboring $J$ levels. Similar effects have been observed at microwave frequencies[18, 20]. Three distinct pathways that can be executed by the action of the second pulse are as follows:

diagonal peak — $|J\rangle\langle J| \xrightarrow{\text{Pulse A}} |J\rangle\langle J+1| \xrightarrow{\text{Pulse B}} |J+1\rangle\langle J+1| \xrightarrow{\text{Pulse B}} |J+1\rangle\langle J|$;

off-diagonal peak — $|J\rangle\langle J| \xrightarrow{\text{Pulse A}} |J\rangle\langle J+1| \xrightarrow{\text{Pulse B}} |J+1\rangle\langle J+1| \xrightarrow{\text{Pulse B}} |J+1\rangle\langle J+2|$;

off-diagonal peak — $|J\rangle\langle J| \xrightarrow{\text{Pulse A}} |J\rangle\langle J+1| \xrightarrow{\text{Pulse B}} |J\rangle\langle J| \xrightarrow{\text{Pulse B}} |J-1\rangle\langle J|$.

In the second pathway, the second field interaction of THz pulse B induces a transition up to $\langle J+2|$ rather than back down to the original $\langle J|$ level. In the third pathway, the first field interaction of THz pulse B returns population to $\langle J|$ rather than promoting population to $\langle J+1|$. These spectral peaks are located at $J$-resolved positions as shown in the 2D $J$-number map plotted as a function of initial and final rotational level $J_i$ and $J_f$ (related to frequencies variables by $\nu = 2Bc(J_i+1)$ and $f = 2Bc(J_f+1)$) in Figs. 4c and 4d. Exemplar pathways for each signal type are elaborated in the double-sided Feynman diagrams shown Fig. S9 in the Supplementary Information.

In addition to the third-order signals, we observed spectral peaks including 2-quantum rephasing (2Q-R) signals and off-diagonal NR and R signals coupling $J$ and $J\pm 2$ levels that are due to five THz field interactions with the dipoles. Examples (described by double-sided Feynman diagrams shown in Fig. S9 in the Supplementary Information) are as follows. The 2Q-R peaks arise firstly through two field interactions by pulse A to create 2QCs $|J\rangle\langle J+2|$ which are then followed by three field interactions with pulse B, rather than one interaction as in the third-order 2Q signals. The first two interactions with pulse B create a population $|J+2\rangle\langle J+2|$ and the third induces a rephasing 1QC $|J+3\rangle\langle J+2|$ or $|J+2\rangle\langle J+1|$ which radiates during $t$. The 2Q-R signals were observed directly in the time domain (see Supplementary Information) and they give rise to peaks along $f=2\nu$ in the R quadrant shown in Fig. 4d. For the fifth-order NR and R signals shown in Figs. 4c and 4d, THz pulse A induces 1QCs $|J\rangle\langle J+1|$ evolving during $\tau$, and pulse B promotes them via four field-dipole interactions to fifth-order 1QCs with final rotational level, $J_f$, 2 quanta away from the initial level $J_i$ (namely, $|J_f - J_i| = 2$) at $|J+3\rangle\langle J+2|$ (R) or $|J+2\rangle\langle J+3|$ (NR). The fifth-order 1QCs then radiate signals correlated to the 1QCs induced during $\tau$. The experimental spectra are in good agreement with simulated 2D spectra shown in Fig. 5 which are generated from double Fourier transformation of the simulated 2D time-domain nonlinear orientation response, $d\cos\langle\theta\rangle_{\mathrm{NL}}/dt$ with respect to $t$ and $\tau$. The simulated spectra capture all the third- and fifth-order signals discussed above.

**Discussion**

We have shown that the main features of the R, NR, 2Q, PP, and 2Q-R signals are in good agreement with our expectations for third- and fifth-order interactions between the incident THz pulse pair and the molecular dipoles and with our numerical simulations based on those expectations. There are weaker sets of features that are not readily explained on the same basis. The experimental R and NR spectra, in addition to the diagonal and off-diagonal peaks attributed to third- and fifth-order field-dipole interactions of the two incident pulses, include weaker, farther off-diagonal peaks that are apparent in the 2D $J$-number plots showing correlations between initial levels $J_i$ and final levels $J_f = J_i + n$ ($n = 3, 4, 5$ etc) in Figs. 4c and 4d. These features are also displayed in 1D line-out plots in Supplementary Information Fig. S5. The relative strengths of the off-diagonal features do not vary substantially with gas pressure (see Fig. S5) or incident field strength, indicating respectively that they are not due to intermolecular dipole-dipole interactions or to high-order interactions with the field. We suggest that the distinct features may be explained by the collective polarization formed by the oriented dipoles during each revival and the resulting nonlinear signal emission, i.e. by superradiance. We have shown previously that this unique type of superradiance results in a series of sudden reductions of the induced rotational populations in every $J$ level, which must occur in order to provide the energy carried away by each FID burst[24]. Coupling among all of the states involved in superradiant emission may be highlighted in the 2D spectrum which displays correlations among their transitions explicitly. Simulations involving the superradiance from the collective polarization are under way, and are beyond the scope of the current study.

In conclusion, we have measured the 2D rotational spectra of molecular rotors using two time-delayed single-cycle THz pulses. We have observed all of the third-order signal contributions expected and some fifth-order signals. The approach we have demonstrated permits direct measurement of rotational dephasing and population relaxation dynamics and spectral correlations that are hidden in linear rotational spectra. With a third THz pulse and a second controlled inter-pulse time delay[30], we hope to observe the time-dependent growth of off-diagonal spectral peaks that could reveal specific relaxation pathways as in other 2D spectroscopies[31]. Independent control of the THz pulse polarizations could reveal the dynamics of relaxation among the $M$ sublevels of the rotational $J$ levels. The approach could provide information that is complementary to conventional microwave rotational spectroscopies[32] and laser centrifuge experiments[33-35].

**Methods**

**Experimental setup.** The laser we used was a Ti:Sapphire amplifier system (Coherent Inc.) operating at 1 kHz repetition rate with an output power of 4 W delivering pulses at 800 nm with 100 fs duration. 95% of the laser output was split equally into two optical paths with a controlled time delay between them. The two delayed optical pulses were recombined in a lithium niobate crystal to generate two time-delayed collinearly propagating THz pulses by optical rectification utilizing the tilted-pulse-front technique[36, 37]. The generated THz pulses were collimated and focused by a pair of 90-degree off-axis parabolic mirrors, resulting in peak field strengths of 400 kV/cm in each pulse reaching the sample gas cell. The transmitted THz fields and THz emission signals from molecular orientation were collimated and refocused by another pair of parabolic mirrors. The remaining 5% of the laser output was used to detect the electric field profile of the THz signals by EO sampling in a 2 mm ZnTe crystal at the focus of the last parabolic mirror.

**Differential chopping detection.** We used two optical choppers at frequencies of 500 Hz and 250 Hz to modulate the two variably delayed optical pulses for THz generation. In successive laser shots, the choppers allowed generation of both THz pulses A and B; A only; B only; and neither pulse. The choppers and laser were synced to a DAQ card (National Instruments) which measured the signal from EO sampling[38]. With background noise subtraction and averaging over 50 laser shots for each data point, the sensitivity of the experiment was in excess of $10^{-3}$. The total data acquisition time for a complete 2D spectrum was typically about 7 days. Real-time measurement of the THz signal field[39] could reduce the acquisition time by alleviating the need for scanning of the EO sampling time variable $t$.

**Sample details.** $CH_3CN$ is a prolate symmetric top molecule with two identical moments of inertia along axes which are perpendicular and much larger than the moment of inertia of the dipole axis. The THz-dipole interaction induces rotation of the dipole axis but does not couple to the rotations about the dipole axis (the selection rule for the transition is $\Delta K = 0$[40]). Consequently, $CH_3CN$ can be approximated as a linear molecule according to the quantum mechanical rigid rotor model[41].

To prepare the $CH_3CN$ gas, liquid $CH_3CN$ was frozen in a storage vessel using liquid nitrogen. The storage vessel, a transfer line, and the gas cell were evacuated. The storage vessel containing $CH_3CN$ was then warmed to room temperature. The vaporized $CH_3CN$ was allowed to diffuse into the gas cell with a path length of 1.25 cm until the pressure in the cell equilibrated with that in the vessel and reached 70 torr, the vapor pressure of $CH_3CN$ at room temperature.

**Acknowledgements.** This work was supported in part by Office of Naval Research Grant No. N00014-13-1-0509 and DURIP grant No. N00014-15-1-2879, National Science Foundation Grant No. CHE-1111557, and the Samsung GRO program. We thank Susan L. Dexheimer, Daniel Kleppner, Robert W. Field, Stephen L. Coy, Colby P. Steiner and Samuel W. Teitelbaum for stimulating discussions.

**Author contributions.** K.A.N. and S.F. conceived the experimental idea. H.Y.H. and J.L. designed the experiment. J.L. collected and analyzed the data. Y.Z. performed numerical simulation and assisted with experiment and analysis. H.Y.H contributed to data analysis and modeling. B.K.O.-O. wrote the LabView code for data acquisition. All authors contributed to the understanding of the underlying light-matter interactions and to writing the manuscript.

**Author Information.** The authors declare no competing financial interests.

# References

1   Ramasesha, K., De Marco, L., Mandal, A. and Tokmakoff, A. Water vibrations have strongly mixed intra- and intermolecular character. *Nature Chem.* **5,** 935-940 (2013).

2   Baiz, C.R., Reppert, M. and Tokmakoff, A. Amide I two-dimensional infrared spectroscopy: methods for visualizing the vibrational structure of large proteins, *J. Phys. Chem. A* **117,** 5955-5961 (2013).

3   Krummel, A.T., Mukherjee, P. and Zanni, M.T. Inter and intrastrand vibrational coupling in DNA studied with heterodyned 2D-IR Spectroscopy. *J. Phys. Chem. B* **107,** pp 9165–9169 (2003).

4   Turner, D.B. and Nelson, K.A. Coherent measurements of high-order electronic correlations in quantum wells. *Nature* **466,** 1089-1092 (2010).

5   Eisele, D.M. *et al*. Robust excitons inhabit soft supramolecular nanotubes. *Proc. Natl. Acad. Sci. USA* **111,** E3367-E3375 (2014).

6   Kampfrath, T., Tanaka, K. and Nelson, K. A. Resonant and nonresonant control over matter and light by intense terahertz transients. *Nature Photon.* **7**, 680 -690 (2013).

7   Somma, C., Reimann, K., Flytzanis, C., Elsaesser, T. and Woerner, M. High-field terahertz bulk photovoltaic effect in lithium niobate. *Phys. Rev. Lett.* **112,** 146602 (2014).

8   Frostig, H., Bayer, T., Dudovich, N., Elda, Y.C. and Silberberg, Y. Single-beam spectrally controlled two-dimensional Raman spectroscopy. *Nature Photon.* **9,** 339-343 (2015).

9   Savolainen, J., Ahmed, S. and Hamm, P. Two-dimensional Raman-terahertz spectroscopy of water. *Proc. Natl. Acad. Sci. USA* **110,** 20402-20407 (2013).

10  Itatani, J. *et al.* Tomographic imaging of molecular orbitals. *Nature* **432,** 867-871 (2004).

11  Vozzi, C. *et al.* Generalized molecular orbital tomography. *Nature Phys.* **7,** 822-826 (2011).

12  Ghafur, O. *et al*. Impulsive orientation and alignment of quantum-state-selected NO molecules. *Nature Phys.* **5,** 289-293 (2009).

13  Fleischer, S., Averbukh, I.Sh. and Prior, Y. Selective alignment of molecular spin isomers. *Phys. Rev. Lett.* **99,** 093002 (2007).

14  Karras, G. *et al*. Orientation and alignment echoes. *Phys. Rev. Lett.* **114,** 153601 (2015).

15  Jiang, H. *et al*. Alignment structures of rotational wavepacket created by two strong femtosecond laser pulses. *Opt. Express* **18**, 8990-8997 (2010).

16  Andrews, D.A., Baker, J.G., Blundell, B.G. and Petty, G.C. Spectroscopic applications of three-level microwave double resonance. *J. Mol. Struct.* **97,** 271 (1983).

17  Stahl, W., Fliege, E. and Dreizler, H. Two-dimensional microwave Fourier transform spectroscopy. *Z. Naturforsch., A* **39,** 858 (1984).

18  Vogelsanger, B. and Bauder, A. Two-dimensional microwave Fourier transform spectroscopy. *J. Chem. Phys.* **92,** 4101 (1990).

19  Twagirayezu, S., Clasp, T.N., Perry, D.S., Neil, J.L., Muckle, M.T. and Pate, B.H. Vibrational coupling pathways in methanol as revealed by coherence-converted population transfer Fourier transform microwave infrared double-resonance spectroscopy. *J. Phys. Chem. A* **114 (25),** 6818–6828 (2010).

20  Wilcox, D.S., Hotopp, K.M. and Dian, B.C. Two-dimensional chirped-pulse Fourier transform microwave spectroscopy. *J. Chem. Phys. A* **115,** 8895 (2011).

21  Martin-Drumel, M.-A., McCarthy, M.C., Patterson, D., McGuire, B.A. and Crabtree, K.N. Automated microwave double resonance spectroscopy: A tool to identify and characterize chemical compounds. *J. Chem. Phys.* **144,** 124202 (2016).

22  Fleischer, S., Zhou. Y., Field, R.W. and Nelson, K.A. Molecular orientation and alignment by intense single-cycle THz pulses. *Phys. Rev. Lett.* **107,** 163603 (2011).

23  Fleischer, S., Field, R.W. and Nelson, K.A. Commensurate two-quantum coherences induced by time-delayed THz fields. *Phys. Rev. Lett.* **109**, 123603 (2012).

24  Fleischer, S. Field, R.W. and Nelson, K.A. From populations to coherences and back again: a new insight about rotating dipoles. arXiv:1405.7025v4 [physics.atom-ph].

25  Hwang, H.Y. *et al*. A review of non-linear terahertz spectroscopy with ultrashort tabletop-laser pulses. *J. Mod. Opt.* **62,** 1447–1479 (2015).

26  Harde, H., Keiding, S. and Grischkowsky, D. THz commensurate echoes: Periodic rephasing of molecular transitions in free-induction decay. *Phys. Rev. Lett.* **66,** 1834-1837 (1991).

27  Siegman, A.E. Lasers. (*University Science Books*1986).

28  Nahata, A., Auston, D.H., Heinz, T.F. and Wu, C. Coherent detection of freely propagating terahertz radiation by electro-optic sampling. *Appl. Phys. Lett.* **68**, 150-152 (1996).

29  Woerner, M., Kuehn, W., Bowlan, P., Reimann, K. and Elsaesser, T. Ultrafast two-dimensional terahertz spectroscopy of elementary excitations in solids. *New J. Phys.* **15,** 025039 (2013).

30  Somma, C., Folpini, G., Reimann, K., Woerner, M. and Elsaesser, T. Two-phonon quantum coherences in indium antimonide studied by nonlinear two-dimensional terahertz spectroscopy. *Phys. Rev. Lett.* **116,** 177401 (2016).

31  Engel, G.S. *et al.* Evidence for wavelike energy transfer through quantum coherence in photosynthetic systems. *Nature* **446,** 782-786 (2007).

32  Park, G.B., Steeves, A.H., Kuyanov-Prozument, K., Neill, J.L. and Field, R.W. Design and evaluation of a pulsed-jet chirped-pulse millimeter wave spectrometer for the 70-102 GHz region. *J. Chem. Phys.* 135, 024202 (2011).

33  Karczmarek, J., Wright, J., Corkum, P. and Ivanov, M. Optical centrifuge for molecules. *Phys. Rev. Lett.* **82,** 3420 (1999).

34  Yuan, L., Teitelbaum, S.W., Robinson, A. and Mullin, A.S. Dynamics of molecules in extreme rotational states. *Proc. Natl. Acad. Sci. USA* **108,** 6872-6877 (2011).

35  Korobenko A., Milner, A.A., Hepburn J.W. and Milner, V. Rotational spectroscopy with an optical centrifuge. *Phys. Chem. Chem. Phys.* **16**, 4071-4076 (2014).

36  Yeh, K.-L., Hoffmann, M.C., Hebling, J. and Nelson K.A. Generation of 10 μJ ultrashort terahertz pulses by optical rectification. *Appl. Phys. Lett.* **90,** 171121 (2007).

37  Hirori, H., Doi, A., Blanchard, F. and Tanaka, K. Single-cycle terahertz pulses with amplitudes exceeding 1 MV/cm generated by optical rectification in LiNbO$_3$. *Appl. Phys. Lett.* **98,** 091106 (2011);

38  Werley, C.A., Teo, S.M. and Nelson, K.A. Pulsed laser noise analysis and pump-probe signal detection with a data acquisition card. *Rev. Sci. Instrum.* **82,** 123108 (2011).

39  Teo, S.M., Ofori-Okai, B.K., Werley, C.A. and Nelson, K.A. Invited Article: Single-shot THz detection techniques optimized for multidimensional THz spectroscopy. *Rev. Sci. Instru*. **86**, 051301 (2015).

40  Khodorkovsky, Y., Kitano, K., Hasegawa, H., Ohshima, Y. and Averbukh, I.Sh. Controlling the sense of molecular rotation: classical versus quantum analysis. *Phys. Rev. A* **83**, 023423 (2011).

41  Townes, C.H. and Schawlow, A.L. Microwave Spectroscopy. (*Dover Publications, Inc., New York* 1975).

**Figures and legends**

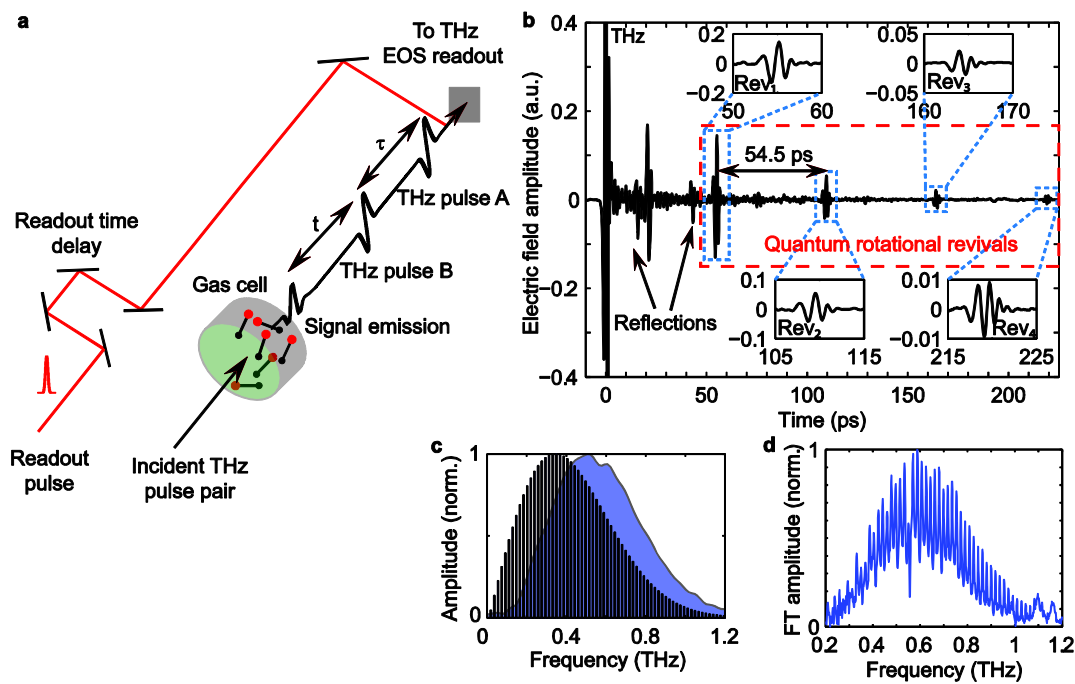

**Figure 1 | Pulse sequence and sample linear response. a,** Two collinearly-propagating THz pulses with a delay $\tau$ are focused into the gas cell with $CH_3CN$. Transmitted THz pulses and induced nonlinear signals are detected by EO sampling (EOS). **b,** Linear FID signal of $CH_3CN$ induced by one THz excitation pulse interacting once with the sample. The signals in the blue dashed boxes are the quantum rotational revivals (enlarged views shown in the insets. $Rev_1$, $Rev_2$ etc. denotes revival 1, 2 etc.). Water vapor in the THz propagation path outside the gas cell and signals owing to double reflection of the THz pulse in the detection crystal and cell windows are observed, but do not affect the nonlinear responses of interest in this work. **c,** Calculated rotational populations of $CH_3CN$ at thermal equilibrium at 300 K as a function of $J\rightarrow J+1$ transition frequency, overlapped with the spectrum of the THz pulses (blue shaded area) used in the experiments. Each bar represents a rotational transition originating from a distinct rotational $J$ level. **d,** Experimental rotational spectrum of $CH_3CN$, Fourier transformed from the FID signal in the red dashed box in **b**. Each sharp peak represents a rotational transition from an initial $J$ level to the final ($J+1$) level. THz absorption by water vapor causes a dip at 0.56 THz in the spectrum.

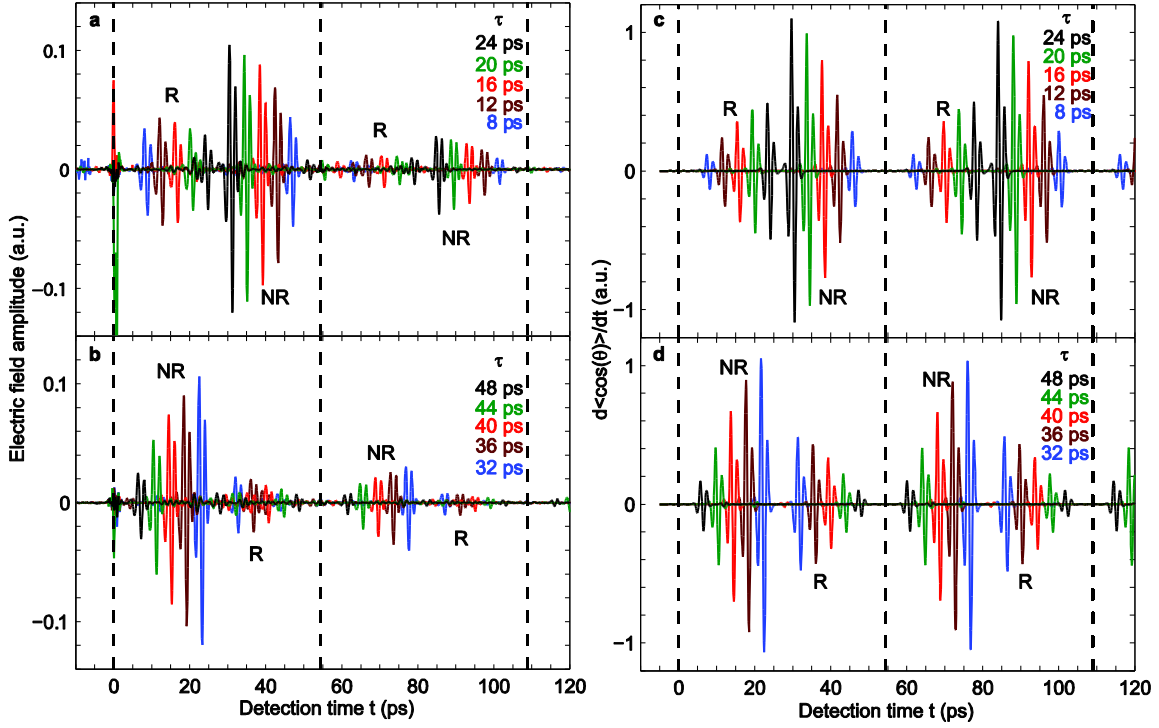

**Figure 2 | THz photon echoes and NR signals in the time domain. a,** Experimental time-domain traces $E_{NL}$ at delays $\tau < T_{rev}/2$ between THz pulse A and B. THz pulse B is fixed at time zero. The rephasing (R) signals appear at $t = \tau + nT_{rev}$ ($n = 0$ and 1 shown) while the non-rephasing (NR) signals appear at $t = -\tau + nT_{rev}$ ($n = 1$ and 2 shown). For $\tau < T_{rev}/2$, each R signal appears earlier than its counterpart NR signal. The vertical dashed lines mark the position of pulse B ($t = 0$) and its first two revivals ($t = T_{rev}$ and $2T_{rev}$), where the pump-probe signals appear. **b,** Experimental time-domain traces $E_{NL}$ at delays $\tau > T_{rev}/2$. Each R signal appears later than the counterpart NR signal. **c** and **d,** Simulated time-domain response of the derivative of the orientation factor, $d\langle\cos(\theta)\rangle/dt$, at delays $\tau < T_{rev}/2$ (**c**) and at delays $\tau > T_{rev}/2$ (**d**). Except for the effects of dephasing which are not accounted for, the simulations show good agreement with the experimental results.

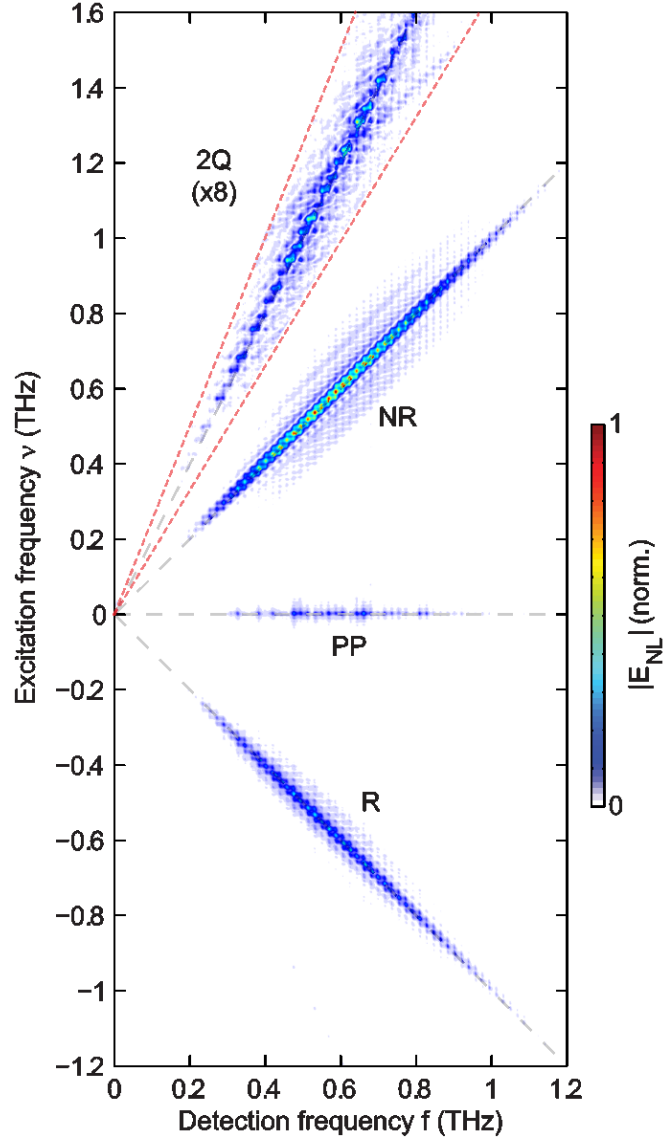

**Figure 3 | 2D THz time-domain rotational spectroscopy.** 2D rotational spectrum obtained by taking the absolute value of the 2D Fourier transformation of the time-domain signal $E_{NL}(\tau,t)$. The light dashed lines are along $\nu = 0$, $\nu = \pm f$, and $\nu = 2f$ respectively. The observed third-order spectral peaks include nonrephasing (NR), rephasing (R), pump-probe (PP) and 2-quantum (2Q, magnified x8 inside the red dashed area) signals. The spectrum is normalized and plotted according to the colormap shown.

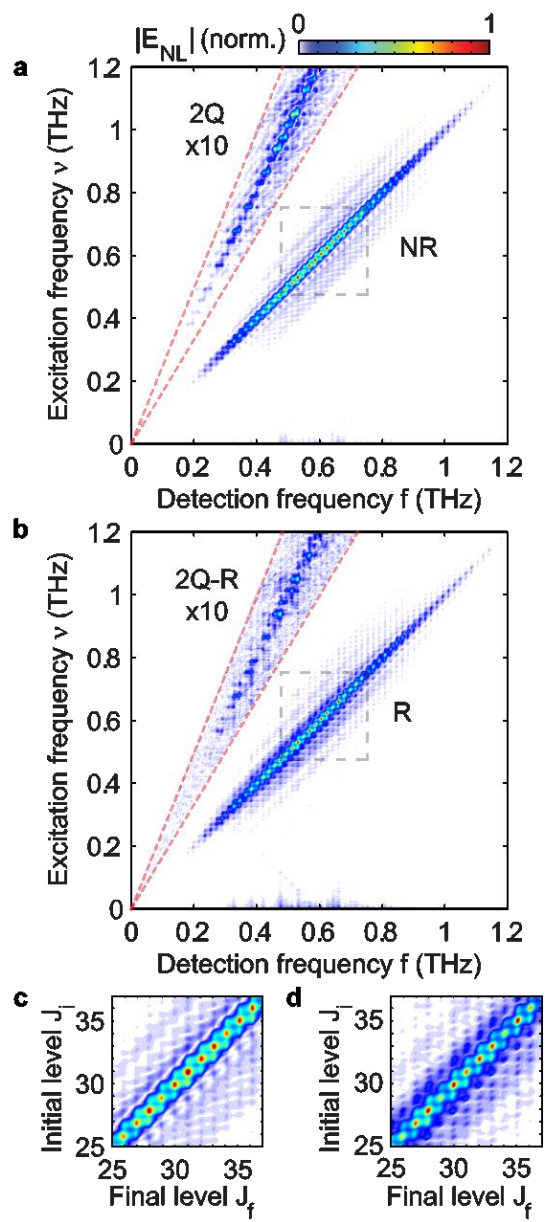

**Figure 4 | Experimental 2D THz rotational spectra. a** and **b,** NR (**a**) and R (**b**, excitation frequency shown as positive) quadrants of the 2D rotational spectrum of CH$_3$CN. Spectral amplitudes inside the red dashed area are magnified 10x to bring out the 2Q (**a**) and 2-quantum rephasing (2Q-R, **b**) signals. The dashed boxes cover rotational transitions from $f_{25, 26}$ to $f_{37, 38}$. **c** and **d,** enlarged views of the spectra within the dashed boxes in the NR quadrant (**a**) and R quadrant (**b**) as functions of initial and final *J* quantum numbers along the vertical and horizontal axes respectively. Third- and fifth-order off-diagonal peaks are separated from the diagonal peaks at *J*-resolved positions. All the spectra are normalized and plotted based on the colormap shown.

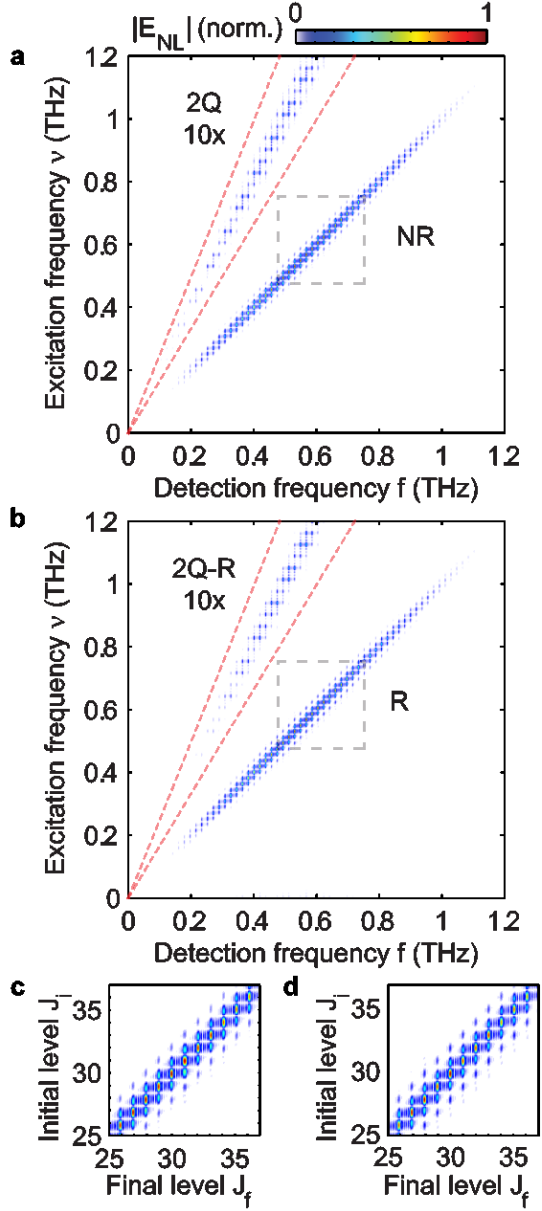

**Figure 5 | Simulated 2D THz rotational spectra. a** and **b,** NR (**a**) and R (**b**) quadrants of the 2D magnitude spectra obtained from numerical Fourier transformation of the simulated 2D time-domain rotational response. The excitation frequency of the R quadrant is made positive. Spectral amplitudes inside the red dashed area are magnified 10x to bring out the 2Q (**a**) and 2-quantum rephasing (2Q-R, **b**) signals. The dashed boxes cover rotational transitions from $f_{25, 26}$ to $f_{37, 38}$. **c** and **d,** Enlarged views of the spectral peaks in the dashed boxes in the NR (**a**) and R (**b**) quadrants as functions of rotational quantum number $J$. Our simulation reproduces well the third- and fifth-order $J$-resolved spectral peaks observed experimentally. All the spectra are normalized and plotted based on the colormap shown.

# Nonlinear two-dimensional terahertz photon echo and rotational spectroscopy in the gas phase

Jian Lu[1], Yaqing Zhang[1], Harold Y. Hwang[1], Benjamin K. Ofori-Okai[1], Sharly Fleischer[2] and Keith A. Nelson[1, *]

[1]Department of Chemistry, Massachusetts Institute of Technology, Cambridge, Massachusetts 02139, USA.

[2]Department of Chemical Physics, Tel-Aviv University, Tel Aviv 69978, Israel.

*Email: kanelson@mit.edu

**Supplementary Information**

Differential chopping detection

Because of the collinear experimental geometry, the nonlinear rotational signals induced by two THz pulses are in the same direction as the transmitted THz fields and their individually induced responses. We implemented a differential chopping detection method[1] in order to separate the nonlinear orientation responses induced via interactions with both THz fields from the responses induced by either THz field alone. The nonlinear signal field ($E_{NL}$) can be written as

$$E_{NL}(\tau, t) = E_{AB}(\tau, t) - E_A(\tau, t) - E_B(t)$$

where $E_{AB}$ is the signal electric field with both THz fields on, and $E_A$ and $E_B$ are the signal electric fields induced by THz field A and by THz field B alone. In the experiment, the nonlinear signal is radiated due to the net orientation of the molecular dipole ensemble following multiple THz-dipole interactions induced by both THz pulses.

An example of the data collected is provided in Fig. S1. Figure S1a shows the EO-sampling traces of the electric field obtained when either THz pulse A ($E_A$, red) or THz pulse B ($E_B$, blue) is present. The two large-amplitude features at -6 ps and 0 ps ($\tau = 6$ ps) correspond to THz pulses A and B transmitted through the gas cell with $CH_3CN$. Figure S1b shows the EO-sampling signal when both THz pulses are present at -6 and 0 ps ($E_{AB}$, black). The subsequent signals result from the orientational responses induced in the gas by THz pulse A, THz pulse B and by the joint action of both THz pulses. The nonlinear signals of interest ($E_{NL}$, magenta) are obtained by subtracting $E_A$ and $E_B$ from $E_{AB}$, revealing only the transient responses that are induced by both THz pulse A and B together as shown in Fig. S1b. In the time trace of $E_{NL}$, we observe third-order nonlinear signals, namely the photon echo signal (R) appearing at $t = \tau = 6$ ps, the nonrephasing (NR) signal during the revival of THz pulse A at $t = T_{rev} - \tau = 48.5$ ps, the pump-probe signal (PP) during the revival of THz pulse B at $t = T_{rev} = 54.5$ ps and the 2-quantum signal (2Q) appearing at $t = T_{rev} - 2\tau = 42.5$ ps. In addition, the fifth-order 2-quantum R signal (2Q-R) appears at $t = 2\tau = 12$ ps. As these signals are all associated with the orientation of molecular dipoles, they all show periodic recurrence with the same revival period $T_{rev}$ as the first-order quantum rotational revivals.

Incrementing the inter-pulse delay $\tau$ quasi-continuously and recording the signal field $E_{NL}(\tau,t)$ at each $\tau$, we obtained the 2D time-time plot of $E_{NL}(\tau,t)$ shown in Fig. S2. $R_n$ marks the R signals appearing at $t = \tau + $ $nT_{\text{rev}}(n = 0, 1, 2 ...)$ in the black dashed boxes and NR$_n$ marks the NR signals appearing at $t = -\tau + nT_{\text{rev}}(n = 1, 2, 3 ...)$ in the magenta dash boxes. But for $\tau = \tau' + mT_{\text{rev}} > T_{\text{rev}}$ ($\tau' < T_{\text{rev}}, m = 1, 2, ...$) there are additional echo signals R$_{-m}$ appearing at $= \tau' + (n - m)T_{\text{rev}} > T_{\text{rev}}$ ($m = n, n-1, n-2 ... -1$), earlier than the expression given above. As an example, R$_{-1}$ was recorded and marked in Fig. S2. As during $\tau > T_{\text{rev}}$, the molecular ensemble experiences quantum revivals during which the oriented molecules resemble the state as initially driven by the THz excitation pulse. The phase reversal of the 1QCs hence starts from the revival time and leads to the additional echo signals.

In the time-domain trace in Fig. S2, signals due to double reflections of our laser pulse in a beamsplitter (delayed from the main THz pulse by ~15 ps) and THz pulse reflections in the EO-sampling detection crystal (delayed by ~40 ps from the main pulse) observed were suppressed by the differential chopping detection method. Double reflections of THz pulses in the front window of the gas cell (delayed from the main pulse by ~20 ps) reentered the cell and provided additional field-dipole interactions with CH$_3$CN, which generated nonlinear signals arising from one such interaction as well as one interaction with each of pulses A and B. These stimulated echo and other signals were not analyzed in detail because the additional time delay could not be scanned independently. They have no significant effects on the rotational responses of interest, as we will show in the numerical simulation.

Numerical simulation of molecular rotation dynamics

Numerical calculation based on the density matrix formalism was utilized to simulate the rotational response under the rigid rotor approximation. The Hamiltonian was separated into system Hamiltonian $H_0$ and THz electric field-dipole interaction term $H_1$.

$$H = H_0 + H_1 = \hat{J}^2/2I - \vec{\mu} \cdot \vec{E}_{\text{THz}}(t)$$

$H_0$ which has been solved analytically yields the eigen-energies and basis set for construction of density matrix elements. The initial population is set to by Boltzmann distribution at 300 K, and all the coherence amplitudes are initially zero. For simplicity we only consider $M = 0$ for each $J$ state, with the initial population of the $|J, 0\rangle$ state given by the Boltzmann distribution function including the multiplicity $\rho_{J,0} = Z^{-1}2J(J + 1)e^{-E_J/k_BT}$, where $E_J = 2BcJ(J + 1)$ is the eigen-energy for rotational level $J$, and $Z = \sum_{J=1}^{80} 2J(J + 1)e^{-E_J/k_BT}$ the partition function considering 80 rotational levels populated thermally. The simplification introduced relies on the $\Delta M = 0$ selection rule for the transition induced by our linearly polarized THz field. This simplification significantly reduces the computation time of the numerical simulations and only slightly affects the magnitudes of the basic nonlinear features that are fully captured by the simulation. The interaction potential $H_1$ is used in the time evolution operator, which is Taylor expanded to third order, to calculate the density matrix evolution in the Hilbert space[2, 3]. We do not consider any decoherence or decay processes in the simulation. The ensemble averaged orientation factor $\langle \cos\theta(t) \rangle$ is given by

$$\langle \cos\theta(t) \rangle = Tr[\rho(t)\cos\theta],$$

where $\theta$ is the angle between molecular dipole and the THz electric field polarization direction. The time derivative of $\langle\cos\theta(t)\rangle$ is proportional to the coherent THz field emitted upon orientation of the molecular ensemble.

Based on the above formalism, the nonlinear orientation factor is calculated as the difference between orientation factor with both THz fields on and that with the individual fields on:

$$\langle\cos\theta(\tau,t)\rangle_{\text{NL}} = \langle\cos\theta(\tau,t)\rangle_{\text{AB}} - \langle\cos\theta(\tau,t)\rangle_{\text{A}} - \langle\cos\theta(t)\rangle_{\text{B}}.$$

To simulate the effects of THz double reflection by the cell window, each excitation THz pulse used in simulation is followed by a small THz pulse with a delay of 21 ps and peak amplitude of 4% of the main pulse. Exemplar time-domain traces calculated with a delay $\tau = 6$ ps are shown in Fig. S3. The observed nonlinear signals show good agreement with the experimental data in Fig. S1 and support the origins discussed in the main paper. THz nonlinear signals induced by the THz double reflection from the gas cell window are also observed, but are beyond the scope of current study.

The 2D time-domain trace of the derivative of the nonlinear orientation factor $d\langle\cos\theta(\tau,t)\rangle_{\text{NL}}/dt$, taking into account the THz double reflections and the resulting nonlinear signals, is plotted in Fig. S4 which shows good agreement with the experimental data in Fig. S2 except for the dephasing and decay not considered. Numerical Fourier transformation of the nonlinear orientation factor yields the simulated 2D rotational spectra shown in Fig. 5 in the main text. Third-order NR, R, 2Q, and PP as well as fifth-order 2Q-R and off-diagonal NR and R spectral peaks are all reproduced in the simulation showing good agreement with the experimental results.

Far off-diagonal peaks

Farther off-diagonal peaks are visible in both the NR and R spectra as functions of $J$ quantum numbers as shown in the 2D $J$-number plots at initial levels $J_i = J$ and final levels $J_f = J + n$ ($n = 3, 4, 5$ etc) in Figs. 4c and 4d in the main text. They can be seen more clearly by plotting a one-dimensional slice of the NR or R spectrum for a specific initial $J_i$ value. The 1D spectral slices with $J_i = 31$, i.e. at an excitation frequency corresponding to the $|J = 31\rangle\langle J = 32|$ 1QC, are plotted for the NR and R quadrants at three sample pressures in Fig. S5. On both sides of the diagonal peak, which has the highest amplitude, there are several off-diagonal peaks indicative of rotational coherences separated from the initial 1QC by as many as seven quanta. Figure S5 shows that the relative amplitudes of the off-diagonal peaks do not diminish significantly as the gas pressure is reduced, showing that the peaks do not arise from intermolecular interactions through which rotational energy and angular momentum are exchanged. Measurements with somewhat different THz field amplitudes, varying by a factor of roughly one third, also showed no significant effects on the form of the spectra, indicating that the far off-diagonal peaks do not arise from correspondingly high-order interactions (e.g. 15th-order for seven quanta off the diagonal) between molecular dipoles and the THz field.

Simulations of the 1D spectral slices with $J_i = 31$ are plotted in Fig. S6 for comparison. The off-diagonal peaks separated by more than two quanta from the diagonal are not captured by the simulations. There are very small kinks at distant $J$ positions that may be due to the finite (220 ps) temporal window used for the simulations, beyond which the signals (which were undamped until that time) were set to zero.

2Q and 2Q-R spectra

The spectral peaks involving 2QCs from both simulation and experimental data are separated from the 2D spectra and presented in Figs. S7 and S8 respectively. The signal levels are an order of magnitude lower than the third-order NR, R and PP responses. In the fifth-order nonlinear processes, THz pulse A successively interacts with the dipoles twice to create 2QCs between $J$ and $J+2$ states which oscillate at the sum of frequencies of two adjacent rotational transitions. Such 2QCs lead to anisotropy in the molecular polarizability tensor, i.e. to net molecular alignment. THz pulse B then interacts with the created 2QCs three times successively, where the first two interactions create, for example, a fourth-order rotational population at state $J+2$. The fifth field-dipole interaction then creates 1QCs between $J+2$ and $J+3$ leading to the radiation of 2Q-R signals with rephasing 1QCs.

Double-sided Feynman Diagrams

Typical excitation pathways that lead to the diagonal and off-diagonal peaks in the 2D rotational spectra can be described by the double-sided Feynman diagrams listed in Fig. S9. Diagram (i) describes the third-order diagonal peak in the NR quadrant located at $(\nu, f) = (2Bc(J+1), 2Bc(J+1))$, while diagrams (ii) and (iii) describe the third-order off-diagonal peaks located at $(\nu, f) = (2Bc(J+1), 2BcJ)$ and $(\nu, f) = (2Bc(J+1), 2Bc(J+2))$. Other Feynman diagrams can be correlated to the spectral peaks in the 2D spectra accordingly. The time subscripts denote the number of preceding field interactions as in conventional 2D spectroscopies. For example in diagrams (i)-(vi), THz pulse A provides the first field-dipole interaction and pulse B provides the second and third field-dipole interactions. Hence $\tau$ represents $t_1$ and $t$ represents $t_3$ ($t_2 = 0$) in these cases.

**References**

1. Woerner, M., Kuehn, W., Bowlan, P., Reimann, K.and Elsaesser, T. Ultrafast two-dimensional terahertz spectroscopy of elementary excitations in solids. *New J. Phys.* **15,** 025039 (2013).
2. Tannor, D. J. Introduction to Quantum Mechanics: A Time-Dependent Perspective (*University Science Books. 2007*).
3. Arfken, G. B. and Weber, H. J. Mathematical Methods for Physicists (*Sixth edition, Elsevier Inc. 2005*).

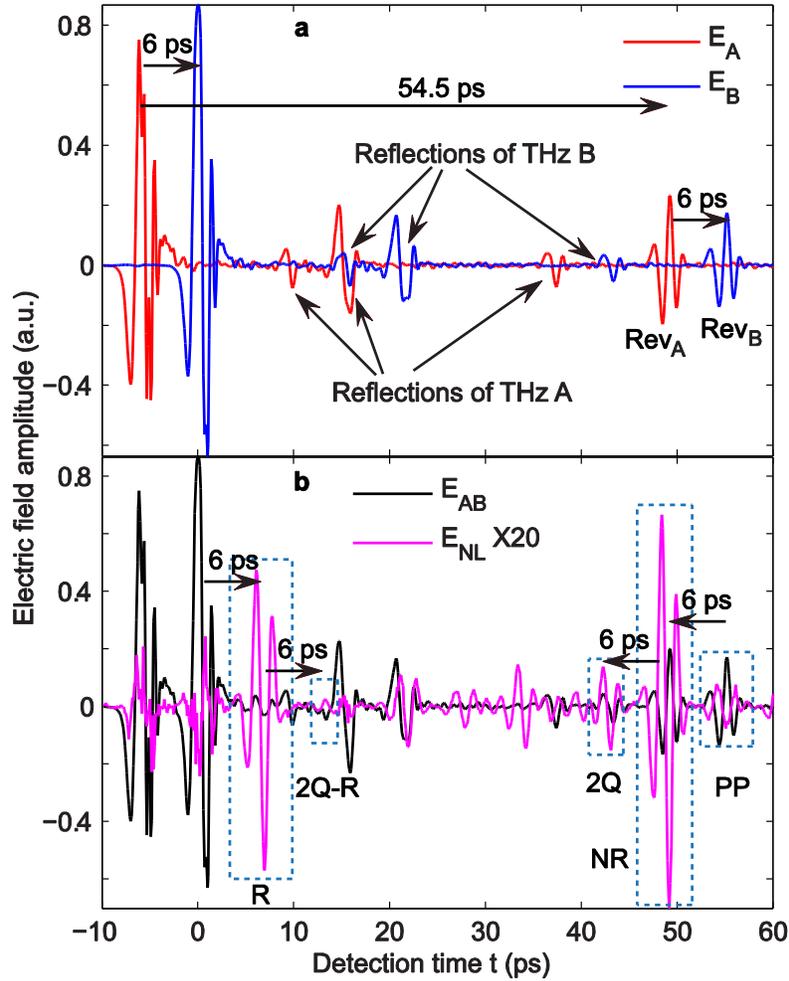

**Figure S1 | Differential chopping detection. a,** The time-domain response with either THz pulse A (red, $E_A$) or B (blue, $E_B$) present. THz pulse B is at 0 ps while THz pulse A is at -6 ps, corresponding to $\tau = 6$ ps. The revivals of THz pulse A ($Rev_A$) and B ($Rev_B$) are at 48.5 and 54.5 ps as marked in the figure. The EO sampling measurements (which are based on THz field-induced depolarization of an optical readout pulse) of the transmitted THz pulses are saturated, but the measurements of the weaker nonlinear THz signal fields are not. **b,** The time-domain response with both THz pulses present (black, $E_{AB}$) and the extracted nonlinear signal (magenta, $E_{NL}$ magnified 20x). The photon echo signal (R) appears at 6 ps, and the 2-quantum R signal (2Q-R) appears at 12 ps (i.e. 6 ps after the R signal). The non-rephasing signal (NR) appears at $R_A$, which is 6 ps earlier than $R_B$. The 2Q signal appears 6 ps earlier than the NR signal. The signal at the arrival of $R_B$ is the pump-probe signal (PP). The spikes in $E_{NL}$ at the arrival of each THz pulse are due to the shot-to-shot fluctuation of the laser.

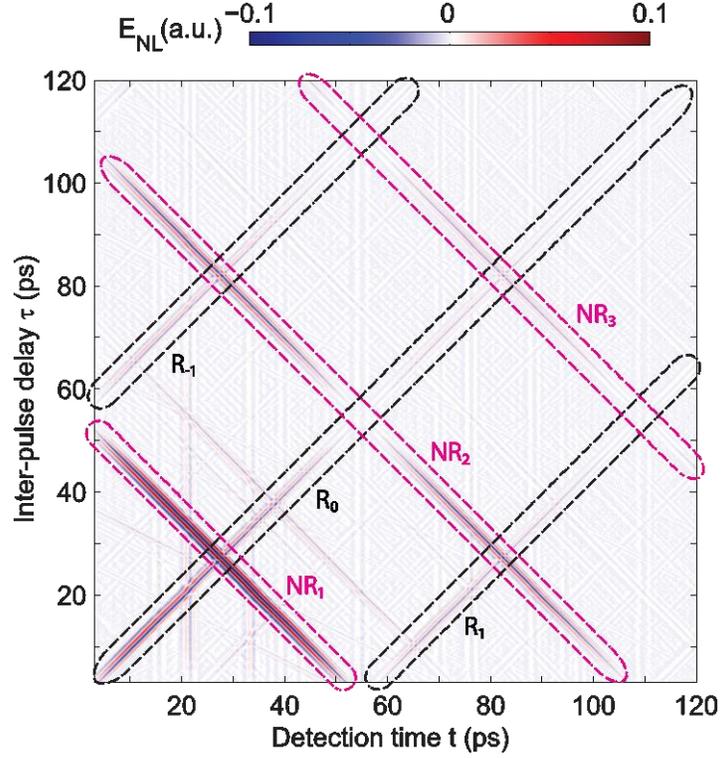

**Figure S2 | Experimental 2D time-domain trace $E_{NL}$.** Experimental 2D time-time plot of $E_{NL}$. THz pulse B is fixed at zero detection time. The delay $\tau$ was incremented from 3 to 120 ps in 200 fs steps, which provided a time window spanning twice the revival period. The detection time $t$ was scanned from 3 to 220 ps (120 to 220 ps not shown) with 100 fs steps. The photon echo (R) signals appear within the black dashed lines and the NR signals within the magenta lines. The subscripts in $R_n$ indicate echo signal timing as $t = \tau + nT_{rev}$. The weak vertical features are pump-probe signals involving various reflections and revivals (the strongest pump-probe signal would appear at detection time $t = 0$). The weak diagonal line from (t = 72 ps, t = 0 ps) to (0 ps, 72 ps) is NR signal involving a reflection. The very weak diagonal line that starts at (38 ps, 0 ps) and declines with half the slope of the NR signals is 2Q signal.

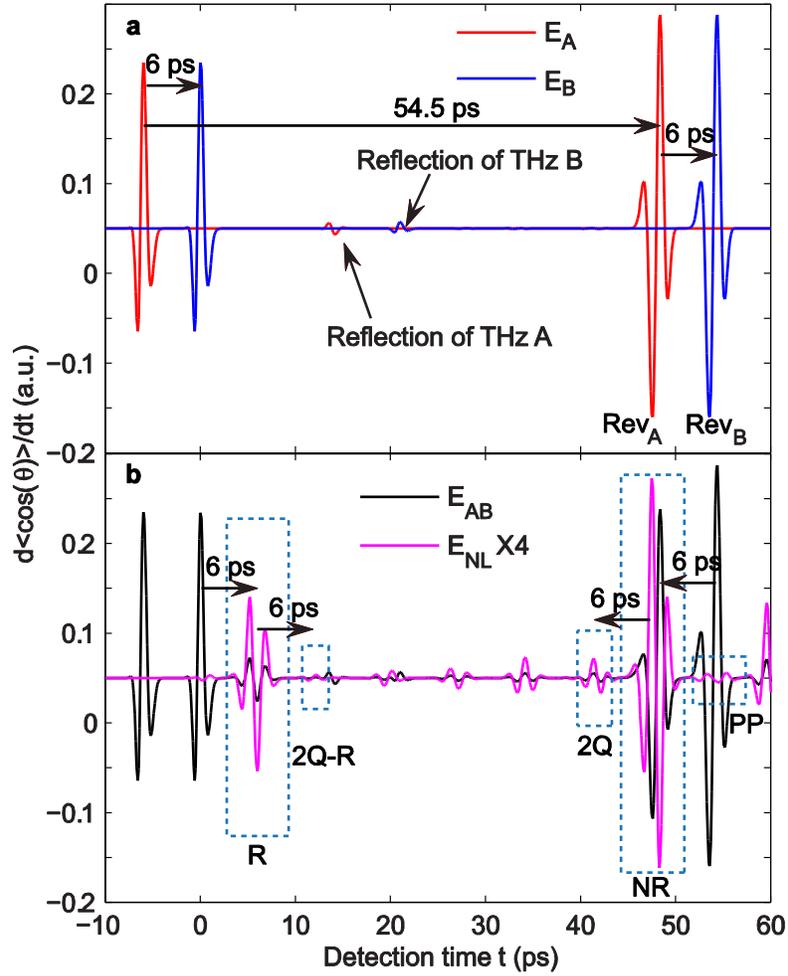

**Figure S3 | Simulated time-domain nonlinear orientation response. a,** The time-domain response of the derivative of orientation factor with either THz pulse A (red, $E_A$) or B (blue, $E_B$) present. THz pulse B is at 0 ps while THz pulse A is at -6 ps. The revivals of THz pulse A ($Rev_A$) and B ($Rev_B$) are at 48.5 and 54.5 ps as marked in the figure. THz double reflections are taken into consideration as marked in the figure. **b,** Time-domain response of the derivative of orientation factor with both THz pulses present (black, $E_{AB}$) and the extracted nonlinear signal (magenta, $E_{NL}$ magnified 4x). The photon echo signal (R) appears at 6 ps, and the 2Q-R signal appears at 12 ps (i.e. 6 ps after the R signal). The NR signal appears at $R_A$, which is 6 ps earlier than $R_B$. The 2Q signal appears 6 ps earlier than the NR signal. The signal at the arrival of $Rev_B$ is the PP signal.

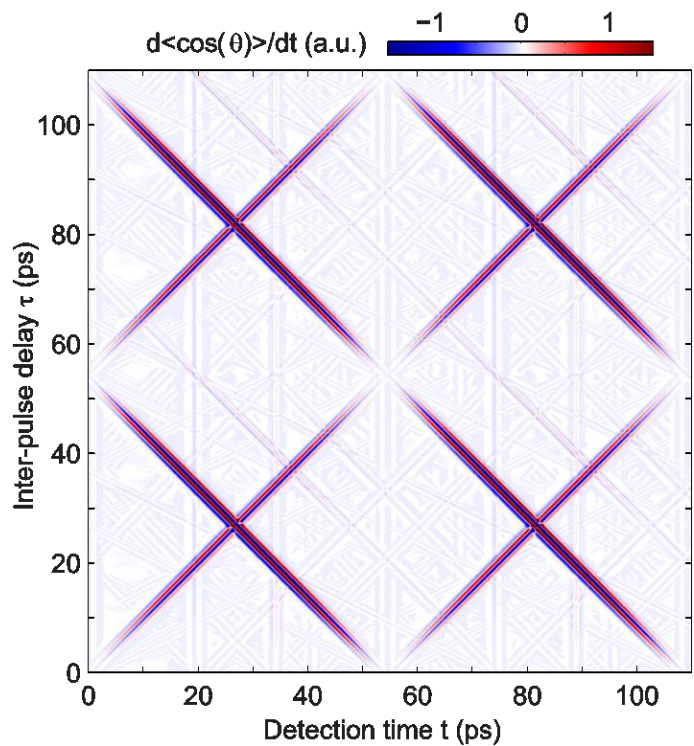

**Figure S4 | 2D time-domain trace of simulated orientation response.** THz double reflections are taken into consideration, resulting in a series of stimulated nonlinear responses. Amplitudes of the derivative of orientation factor $d\langle\cos(\theta)\rangle/dt$ exceeding $\pm 1.5$ are saturated in the colormap for better contrast of weak signals.

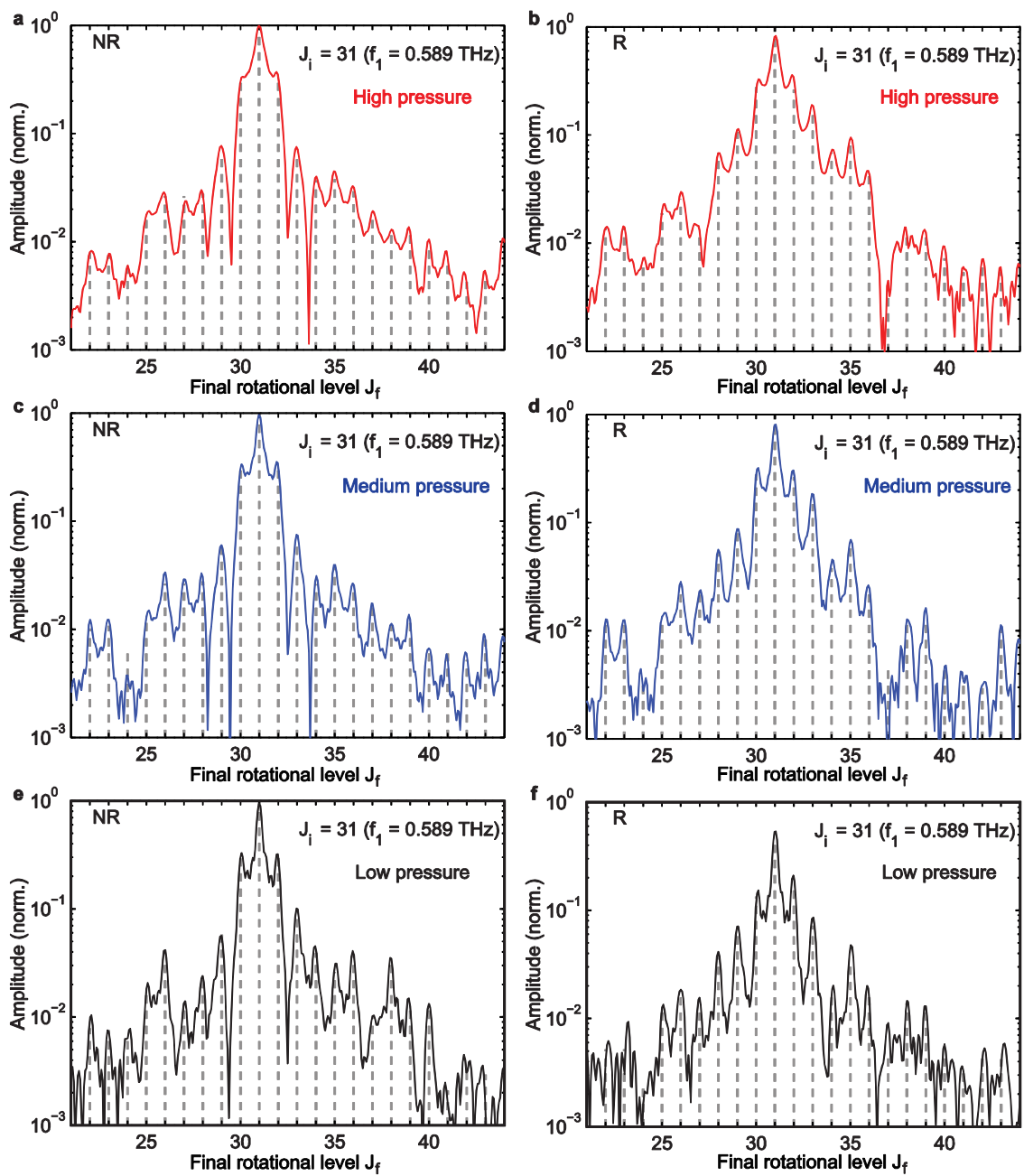

**Figure S5 | Spectral slices from 2D rotational spectra.** 1D spectral slices along excitation frequency $\nu =$ 0.589 THz $= f_{31,32}$ (corresponding to the rotational transition between $J = 31$ and 32) from the NR (**a, c, e**) quadrant and the R quadrant (**b, e, f**) at 70 torr (high pressure, red), 52 torr (medium pressure, blue) and 23 torr (low pressure, black). As pressure decreases, the linewidths of the rotational spectral peaks are narrower due to less collisional broadening. In both plots the vertical dashed lines represent the positions of each rotational transition frequency ranging from $f_{21,22}$ to $f_{44,45}$. The correlation between the rotational coherences as far apart in frequency as $|31\rangle\langle32|$ and $|39\rangle\langle40|$ is seen by the most distant off-diagonal peak above the noise floor.

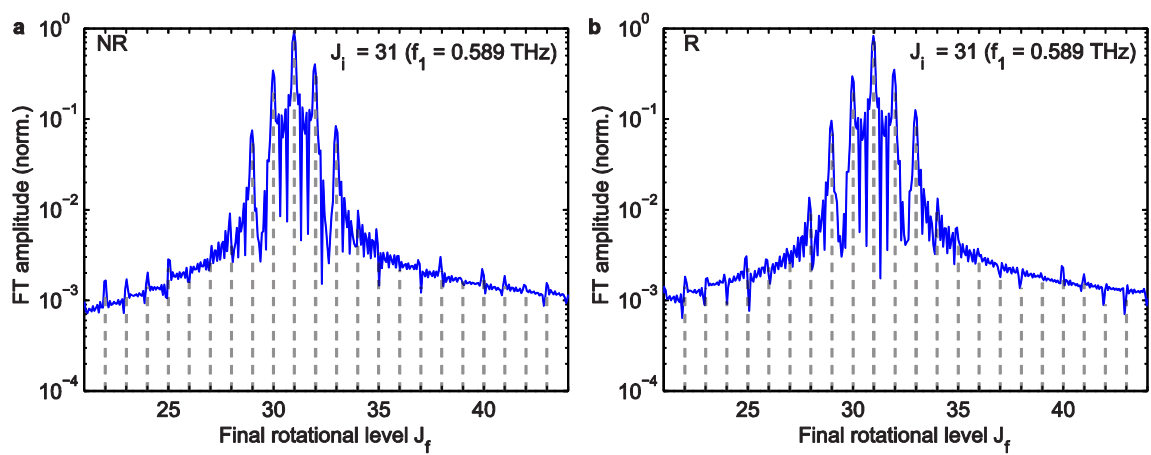

**Figure S6 | Spectral slices from the simulated 2D rotational spectra. a,** 1D spectral slice along $\nu = 0.589$ THz (corresponding to initial $J_i = 31$) in the NR quadrant. The highest amplitude peak at final $J_f = 31$ is the diagonal peak. The two peaks at final $J = 30$ and $32$ are the third-order off-diagonal peaks. The two peaks at final $J = 29$ and $33$ are fifth-order off-diagonal peaks. **b,** 1D spectral slice along $\nu = 0.589$ THz (corresponding to $J_i = 31$) in the R quadrant.

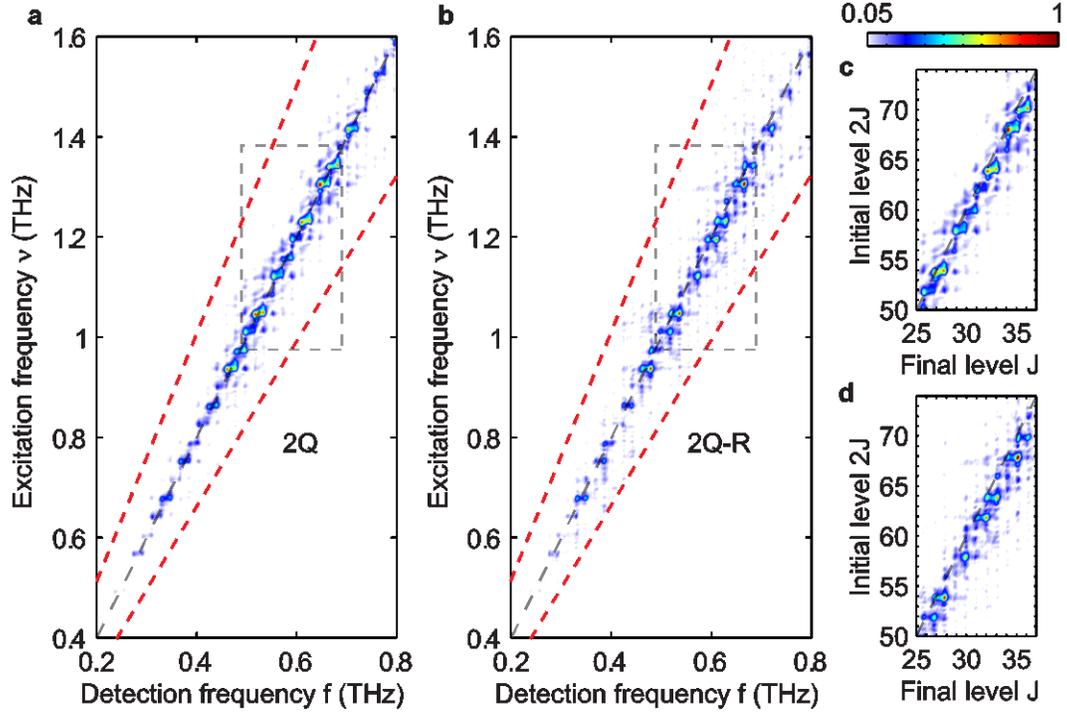

**Figure S7 | Experimental 2Q and 2Q-R spectra. a,** 2Q and **b,** 2Q-R normalized magnitude spectra isolated from the experimental 2D spectra shown in the main text. The amplitudes outside the red dashed lines are set to zero. The spectral peaks between the red dashed lines are normalized and plotted according to the color coding shown. **c** and **d,** 2D *J*-quantum-number plots, i.e. magnified views of the spectra in the dashed boxes with the initial and final *J* levels for the transitions indicated along the vertical and horizontal axes respectively. The strongest diagonal peaks are at $f = 2Bc(J+1)$ and $\nu = 2Bc(2J+3)$, while the strongest off-diagonal peaks are at $f = 2Bc(J+2)$ and $\nu = 2Bc(2J+3)$. All the spectra are normalized and plotted according to the colormap shown. The additional features are due to noise.

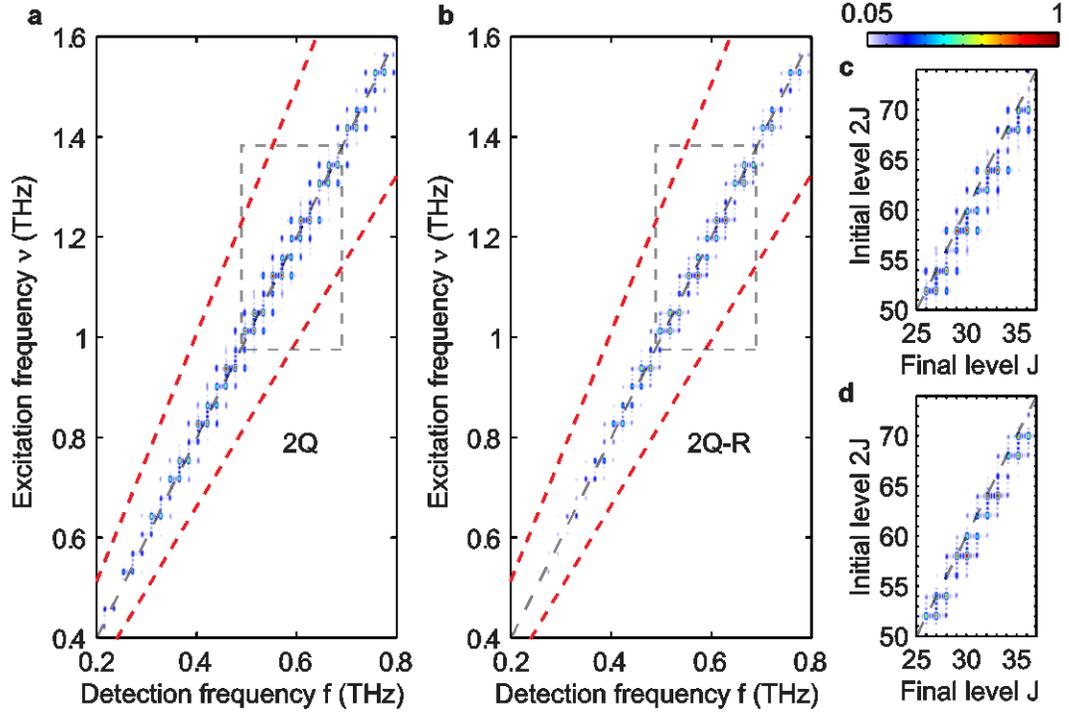

**Figure S8 | Simulated 2Q and 2Q-R spectra. a,** 2Q and **b,** 2Q-R spectra are isolated from the simulated 2D spectra. The amplitudes outside the red dashed lines are set to zero. The spectral peaks between the red dashed lines are normalized and plotted between 0 and 0.6 of the maximum value based on the color coding shown. **c** and **d,** 2D *J*-quantum-number map, i.e. magnified views of the spectra in the dashed boxes. The strongest diagonal peaks are at $f = 2Bc(J+1)$ and $\nu = 2Bc(2J+3)$, while the strongest off-diagonal peaks are at $f = 2Bc(J+2)$ and $\nu = 2Bc(2J+3)$. All the spectra are normalized and plotted according to the colormap shown.

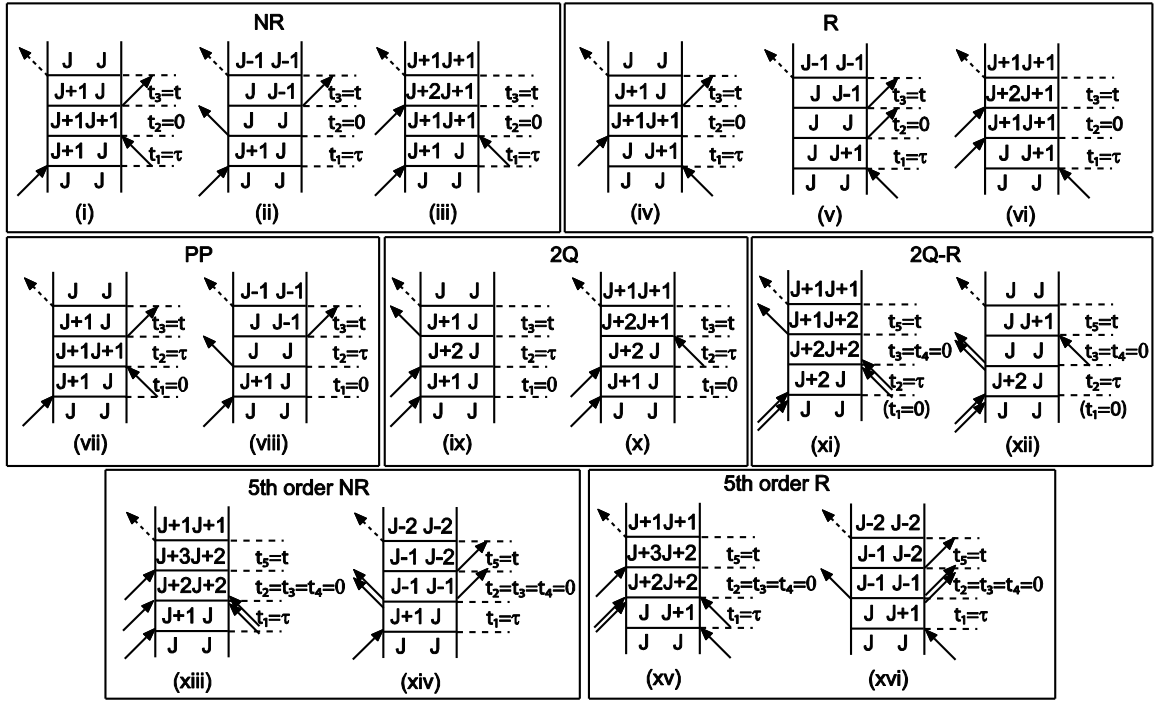

**Figure S9 | Examples of double-sided Feynman diagrams.** Diagram (i)-(iii) describes the third-order diagonal and off-diagonal peaks in the NR quadrant of the 2D spectrum. Diagrams (iv)-(vi) describe the third-order diagonal and off-diagonal peaks in the R quadrant. Diagrams (vii)-(viii) describe two excitation pathways that lead to the PP peaks. Diagrams (ix)-(xii) describe the typical excitation pathways leading to the 2Q and 2Q-R peaks. Diagrams (xii)-(xiv) and (xv)-(xvi) describe the typical excitation pathways leading to the fifth-order NR and R off-diagonal peaks. The time subscripts denote the number of preceding field interactions. The nonlinear signal emission time period $t$ corresponds to $t_3$ in all third-order processes and $t_5$ for the fifth-order processes. The inter-pulse delay time $\tau$ corresponds to $t_1$ for the R and NR signals (for third- and fifth-order signals) and to $t_2$ for the PP, 2Q and 2Q-R signals.